\documentclass[preprint] {aastex}
\usepackage{epstopdf}
\usepackage{gensymb}
\usepackage{natbib}

\shorttitle{Simultaneous Multiwavelength Obs. of Mrk421}

\title{Simultaneous Multiwavelength Observations of Markarian 421 During Outburst}

\author{
V. A. Acciari\altaffilmark{1},
E. Aliu\altaffilmark{17},
T. Aune\altaffilmark{2},
M. Beilicke\altaffilmark{3},
W. Benbow\altaffilmark{1},
M. B{\"o}ttcher\altaffilmark{4},
S. M. Bradbury\altaffilmark{5},
J. H. Buckley\altaffilmark{3},
V. Bugaev\altaffilmark{3},
Y. Butt\altaffilmark{6},
A. Cannon\altaffilmark{7},
O. Celik\altaffilmark{8},
A. Cesarini\altaffilmark{9},
Y. C. Chow\altaffilmark{8},
L. Ciupik\altaffilmark{10},
P. Cogan\altaffilmark{11},
P. Colin\altaffilmark{12},
W. Cui\altaffilmark{13}\altaffilmark{*},
R. Dickherber\altaffilmark{3},
C. Duke\altaffilmark{15},
A. D. Falcone\altaffilmark{30},
S. J. Fegan\altaffilmark{8},
J. P. Finley\altaffilmark{13},
G. Finnegan\altaffilmark{12},
P. Fortin\altaffilmark{16},
L. Fortson\altaffilmark{10},
A. Furniss\altaffilmark{2},
D. Gall\altaffilmark{13}\altaffilmark{*},
G. H. Gillanders\altaffilmark{9},
J. Grube\altaffilmark{7},
R. Guenette\altaffilmark{11},
G. Gyuk\altaffilmark{10},
D. Hanna\altaffilmark{11},
J. Holder\altaffilmark{17},
D. Horan\altaffilmark{18},
C. M. Hui\altaffilmark{12},
T. B. Humensky\altaffilmark{19},
P. Kaaret\altaffilmark{20},
N. Karlsson\altaffilmark{10},
M. Kertzman\altaffilmark{21},
D. Kieda\altaffilmark{12},
J. Kildea\altaffilmark{1},
A. Konopelko\altaffilmark{22},
H. Krawczynski\altaffilmark{3},
F. Krennrich\altaffilmark{23},
M. J. Lang\altaffilmark{9},
S. LeBohec\altaffilmark{12},
G. Maier\altaffilmark{11},
A. McCann\altaffilmark{11},
J. Millis\altaffilmark{24},
P. Moriarty\altaffilmark{25},
R. A. Ong\altaffilmark{8},
A. N. Otte\altaffilmark{2},
D. Pandel\altaffilmark{20},
J. S. Perkins\altaffilmark{1},
A. Pichel\altaffilmark{26},
M. Pohl\altaffilmark{23},
J. Quinn\altaffilmark{7},
K. Ragan\altaffilmark{11},
L. C. Reyes\altaffilmark{27},
P. T. Reynolds\altaffilmark{28},
E. Roache\altaffilmark{1},
H. J. Rose\altaffilmark{5},
M. Schroedter\altaffilmark{23},
G. H. Sembroski\altaffilmark{13},
A. W. Smith\altaffilmark{29},
D. Steele\altaffilmark{10},
S. P. Swordy\altaffilmark{19},
M. Theiling\altaffilmark{1},
J. A. Toner\altaffilmark{9},
A. Varlotta\altaffilmark{13},
S. Vincent\altaffilmark{12},
S. P. Wakely\altaffilmark{19},
J. E. Ward\altaffilmark{7},
T. C. Weekes\altaffilmark{1},
A. Weinstein\altaffilmark{8},
T. Weisgarber\altaffilmark{19},
D. A. Williams\altaffilmark{2},
S. Wissel\altaffilmark{19},
B. Zitzer\altaffilmark{13}\\
(The VERITAS Collaboration),\\
I. de la Calle Perez\altaffilmark{31},
A. Ibarra\altaffilmark{31},
and P. Rodriguez\altaffilmark{31}\\
and\\
H.~Anderhub\altaffilmark{32},
L.~A.~Antonelli\altaffilmark{33},
P.~Antoranz\altaffilmark{34},
M.~Backes\altaffilmark{35},
C.~Baixeras\altaffilmark{36},
S.~Balestra\altaffilmark{34},
J.~A.~Barrio\altaffilmark{34}
% ...
\\
(The MAGIC Collaboration)
}

\altaffiltext{1}{Fred Lawrence Whipple Observatory, Harvard-Smithsonian Center for Astrophysics, Amado, AZ 85645, USA}
\altaffiltext{2}{Santa Cruz Institute for Particle Physics and Department of Physics, University of California, Santa Cruz, CA 95064, USA}
\altaffiltext{3}{Department of Physics, Washington University, St. Louis, MO 63130, USA}
\altaffiltext{4}{Astrophysical Institute, Department of Physics and Astronomy, Ohio University, Athens, OH 45701}
\altaffiltext{5}{School of Physics and Astronomy, University of Leeds, Leeds, LS2 9JT, UK}
\altaffiltext{6}{Harvard-Smithsonian Center for Astrophysics, 60 Garden Street, Cambridge, MA 02138, USA}
\altaffiltext{7}{School of Physics, University College Dublin, Belfield, Dublin 4, Ireland}
\altaffiltext{8}{Department of Physics and Astronomy, University of California, Los Angeles, CA 90095, USA}
\altaffiltext{9}{School of Physics, National University of Ireland, Galway, Ireland}
\altaffiltext{10}{Astronomy Department, Adler Planetarium and Astronomy Museum, Chicago, IL 60605, USA}
\altaffiltext{11}{Physics Department, McGill University, Montreal, QC H3A 2T8, Canada}
\altaffiltext{12}{Physics Department, University of Utah, Salt Lake City, UT 84112, USA}
\altaffiltext{13}{Department of Physics, Purdue University, West Lafayette, IN 47907, USA }
\altaffiltext{15}{Department of Physics, Grinnell College, Grinnell, IA 50112-1690, USA}
\altaffiltext{16}{Department of Physics and Astronomy, Barnard College, Columbia University, NY 10027, USA}
\altaffiltext{17}{Department of Physics and Astronomy and the Bartol Research Institute, University of Delaware, Newark, DE 19716, USA}
\altaffiltext{18}{Laboratoire Leprince-Ringuet, Ecole Polytechnique, CNRS/IN2P3, F-91128 Palaiseau, France}
\altaffiltext{19}{Enrico Fermi Institute, University of Chicago, Chicago, IL 60637, USA}
\altaffiltext{20}{Department of Physics and Astronomy, University of Iowa, Van Allen Hall, Iowa City, IA 52242, USA}
\altaffiltext{21}{Department of Physics and Astronomy, DePauw University, Greencastle, IN 46135-0037, USA}
\altaffiltext{22}{Department of Physics, Pittsburg State University, 1701 South Broadway, Pittsburg, KS 66762, USA}
\altaffiltext{23}{Department of Physics and Astronomy, Iowa State University, Ames, IA 50011, USA}
\altaffiltext{24}{Department of Physics, Anderson University, 1100 East 5th Street, Anderson, IN 46012}
\altaffiltext{25}{Department of Life and Physical Sciences, Galway-Mayo Institute of Technology, Dublin Road, Galway, Ireland}
\altaffiltext{26}{Instituto de Astronomia y Fisica del Espacio, Casilla de Correo 67 - Sucursal 28, (C1428ZAA) Ciudad Aut—noma de Buenos Aires, Argentina}
\altaffiltext{27}{Kavli Institute for Cosmological Physics, University of Chicago, Chicago, IL 60637, USA}
\altaffiltext{28}{Department of Applied Physics and Instrumentation, Cork Institute of Technology, Bishopstown, Cork, Ireland}
\altaffiltext{29}{Argonne National Laboratory, 9700 S. Cass Avenue, Argonne, IL 60439, USA}
\altaffiltext{30}{Pennsylvania State University, Dept. of Astronomy \& Astrophysics, University Park, PA 16802}

\altaffiltext{31}{XMM-Newton-SOC, European Space Astronomy Centre, ESA, 28691 Villanueva de la Ca\~{n}ada (Madrid), Spain}
\altaffiltext{32} {ETH Zurich, CH-8093 Switzerland}
\altaffiltext{33} {INAF National Institute for Astrophysics, I-00136 Rome, Italy}
\altaffiltext{34} {Universidad Complutense, E-28040 Madrid, Spain}
\altaffiltext{35} {Technische Universit\"at Dortmund, D-44221 Dortmund, Germany}
\altaffiltext{36} {Universitat Aut\`onoma de Barcelona, E-08193 Bellaterra, Spain}
% ...
\altaffiltext{*}{Corresponding author:  Daniel Gall, dgall@purdue.edu or Wei Cui, cui@physics.purdue.edu}

\date{}                                           

\begin{document}

\begin{abstract}
\noindent
We report on the results of two coordinated multiwavelength campaigns that focused on the blazar Markarian 421 during its 2006 and 2008 outbursts. These campaigns obtained UV and X-ray data using the {\em XMM-Newton} satellite, while the gamma-ray data were obtained utilizing three imaging atmospheric Cerenkov telescopes, the {\em Whipple} 10m telescope and {\em VERITAS}, both based in Arizona, as well as the {\em MAGIC} telescope, based on La Palma in the Canary Islands.  The coordinated effort between the gamma-ray groups allowed for truly simultaneous data in UV/X-ray/gamma-ray wavelengths during a significant portion of the {\em XMM-Newton} observations.  This simultaneous coverage allowed for a reliable search for correlations between UV, X-ray and gamma-ray variability over the course of the observations.  Investigations of spectral hysteresis and modeling of the spectral energy distributions are also presented.  
\end{abstract}

\section{Introduction}

Blazars, a sub-class of active galactic nuclei (AGN), are some of the most intriguing sources in the high-energy sky.  Their rapid variability and the non-thermal nature of their emission, presenting a continuum across nearly the entire electromagnetic spectrum, imply that the observed photons originated within highly relativistic jets oriented very close to the observer's line of sight \citep{Urry}.  Therefore, blazars are excellent laboratories for studying the physical processes within the jets of AGN.  They were among the first very high energy (VHE; E $>$100 GeV) sources to be detected and today there are 25 known VHE blazars, including flat spectrum radio quasars and BL Lac objects.

Various models have been proposed to account for the broad-band spectral energy distributions (SED)s observed in VHE blazars, which typically display a characteristic double peak when plotted as  $\nu F _\nu$ against $\nu$, with peaks occurring at keV and TeV energies. The models are generally divided into two classes:  leptonic and hadronic.  Both leptonic and hadronic models attribute the peak at keV energies to synchrotron radiation from relativistic electrons (and positrons) within the jet, but they differ on the origin of the TeV peak.  The leptonic models advocate the inverse Compton scattering mechanism, utilizing synchrotron self Compton (SSC) interactions and/or inverse Compton interactions with an external photon field, to explain the VHE emission \citep[e.g.,][]{Marscher,Maraschi,Dermer,Sikora}.  The hadronic models, however, account for the VHE emission by $\pi^0$ or charged pion decay with subsequent synchrotron and/or Compton emission from decay products, or synchrotron radiation from ultra-relativistic protons \citep[e.g.,][]{Mannheim, Aharonian, Pohl}.

Observationally, blazars are known to undergo both major outbursts on long time scales and rapid flares on short time scales, most prominently at keV and TeV energies.  During some outbursts, both of the SED peaks have been observed to shift towards higher energies in a generally correlated manner \citep[e.g.,][]{Blazejowski}.  The correlation of the variabilities at keV and TeV energies (or lack thereof) during such outbursts has aided in refining the emission models.  In addition, rapid, sub-hour flaring is interesting as it provides direct constraints on the size of the emission region.  These rapid flares also present an observational challenge to multiwavelength studies, as truly simultaneous data must be used in order to develop a reliable characterization of the broadband behavior of these objects. 

Markarian 421 (Mrk421; 1101+384), at a redshift of $z=0.031$, was the first blazar, as well as the first extragalactic source, to be detected at TeV energies \citep{Punch} and has since remained one of the most active VHE blazars.  Its SED has peaks at keV and TeV energies, and it has been known to demonstrate rapid, sub-hour, flaring behavior at these energies during the course of an outburst \citep[e.g.,][]{Cui,Gaidos} indicating very compact emission regions.  Many extensive multiwavelength campaigns studying Markarian 421 have been undertaken at keV and TeV energies during outbursts, but the degree of simultaneity of the multiwavelength coverage varies and is often not adequate to account for the most rapid variability of the source.  Due to these considerations, one must exercise caution in the interpretation of some of the results from these campaigns.

In this work, we report on results from a target-of-opportunity (ToO) program on VHE blazars that makes use of a unique combination of capabilities provided by the {\em XMM-Newton} satellite and several ground-based imaging atmospheric Cerenkov Telescopes (IACT)s to obtain truly simultaneous coverage in the optical/UV, X-ray and VHE bands.  Unlike other satellites frequently used for multiwavelength campaigns, {\em XMM-Newton's} highly elliptical orbit allows for long observations that are not frequently interrupted by earth occultation.  In addition to its X-ray instruments, {\em XMM-Newton} carries an optical/UV telescope co-aligned with the X-ray telescopes and can thus provide simultaneous coverage in the optical/UV band.  While our primary emphasis is on the X-ray and VHE bands, the optical/UV coverage helps to constrain the overall SED shape.

\section{Observations and Data Reduction}

Our ToO program was first triggered in April 2006 by a major outburst from Markarian 421 as detected by regular monitoring of the VHE band by the {\em Whipple} 10m telescope.  Because of {\em XMM-Newton} visibility constraints, the coordinated multiwavelength observations did not take place until after the peak of the outburst as indicated by the overall monitoring campaign.  The program was triggered again in May 2008 by another major outburst from Markarian 421 detected in the VHE band.  However, we once again captured only the decaying portion of the outburst.  However, taken together, the two campaigns have produced a significant amount of simultaneous optical/UV, X-ray and VHE data on the source.

\subsection{XMM-Newton Observations}

The X-ray and optical/UV observations  were taken by the {\em XMM-Newton} satellite's EPIC-pn (EPN) detector \citep{Struder}, covering a spectral range of approximately 0.5 - 10 keV, and the Optical Monitor \citep[OM;][]{Mason}, capable of covering the range between 170 and 650 nm (7.3 eV and 1.9 eV).  The metal oxide semiconductor (MOS) detectors were also operated during some of the observations, but the data were not used here.  During both observations, the EPN was operated in fast timing mode to minimize photon pileup at the expense of imaging capability along the direction of event readout.  The thin optical filter was in place for both observations.  The 2006 and 2008 observations produced EPN exposures of approximately 42 ks and 43 ks, respectively.  Due to the brightness of the source, during the first half of the 2008 observations both the EPN and MOS detectors entered into counting mode, resulting in a loss of frames.  In an attempt to avoid further frame losses, the MOS detectors were shut down to provide more bandwidth for telemetry to the EPN detector.  As a result of the detector being in counting mode, the first half of the EPN observation suffers from telemetry gaps which can be accounted for and corrected as explained below.

In parallel with the EPN observations, a series of exposures was taken utilizing the OM.  The 2006 and 2008 observations produced 15 and 10 exposures, respectively, with total exposure times of 32.5 ks and 22.0 ks.  For all exposures, the OM was in imaging mode with the UVM2 filter (200 - 300 nm) in place.  Table \ref{table:xmmobs} summarizes these observations.

The 2006 EPN data were initially processed using XMMSAS v7.0 \citep{Gabriel}.  Standard XMM-Newton data analysis procedures were followed to filter and reduce the data, generate the various lightcurves, extract the source and background spectrum, and generate the RMF and ARF files for subsequent spectral analyses.  In summary, the event list was examined to check for periods of soft proton flaring, where background event rates are higher than 0.4 cts/sec (for events with energies E $>$ 10 keV).  In addition, ``bad events" were removed including events which were close to CCD gaps or bad pixels, and only single and double photon events were included (PATTERN$\leq$4).  A one dimensional histogram of counts was produced to determine the position of the source.  Based on the histogram, we chose columns 30-45 (in RAWX) for the source region and columns 1-15 and 57-64 for the background region.  These regions were used to extract source and background spectra.

The 2008 EPN data were processed similarly, using the same filtering criteria, but with XMMSAS v8.0.  In this case the source region was determined to be in columns 31-45 and the background regions were taken to be columns 1-19 and 57-63.  It should be noted that  a new SAS task, {\em epiclccorr}, was used to correct for the effects of deadtime, telemetry saturation, and frame dropouts in producing the light curves from the 2006 and 2008 data.

\par The OM data were analyzed with the standard SAS task {\it omichain},
which produces images and source lists for all OM exposures present in any
given observation. The 2006 OM observations presented here are a combination of 15
exposures with 2000~s and 2500~s integration times while the 2008 data comes from 10 exposures each with a duration of 2200 s.

\subsection{VHE Gamma-Ray Observations}
Three ground-based IACT facilities were utilized to observe the source in the VHE band during the {\em XMM-Newton} observations.  These IACTs detect gamma-rays by imaging the flashes of Cerenkov light
emitted by gamma-ray induced electromagnetic showers within the atmosphere.  The {\em Major Atmospheric Gamma Imaging Cerenkov} (MAGIC) telescope provided 14 ks of coverage during the first part of the 2006 {\em XMM-Newton} observation, and the {\em Whipple} 10m telescope provided 12 ks of coverage during the latter part of the 2006 observation.  The geographic separation of the experiments allowed for extended, simultaneous coverage during the {\em XMM-Newton} observation.  During the 2008 {\em XMM-Newton} observation, only the {\em Very Energetic Radiation Imaging Telescope Array System} (VERITAS) was used, providing 9 ks of VHE coverage.

\subsubsection{Whipple Data} \label{sec:whipple}

The {\em Whipple} 10m IACT \citep{Kildea} focuses Cerenkov light on a camera composed of 499 photo multiplier tubes (PMTs).  The energy threshold for the {\em Whipple} 10m is near 400 GeV.  There are two modes used for taking data.  During the 2006 observation, two hours of data were taken in TRACKING mode, where the source is centered in the camera, and two hours of data were taken in ON/OFF mode, where half of the data runs are offset from the source by 30 minutes in right ascension to provide an independent measurement of background events (primarily cosmic-rays).  The runs each had a duration of 28 minutes, and the total time on source for the {\em Whipple} 10m telescope was 3 hours on 2006 April 30 from 4:37 UT to 8:05 UT.  The observations were taken with a source elevation range of 45$\degree$ -- 79$\degree$, with an average elevation of 64$\degree$ and during favorable weather conditions.  

{\em Whipple} data reduction involves two stages, image cleaning and image parameterization.  In the cleaning stage, the data is flat-fielded using a run during which a nitrogen arc lamp is pulsed to illuminate the camera's pixels.  Any differences between night-sky background between ON and OFF runs is also accounted for at this stage.  In addition, pixels that make up shower images are selected and the remaining background pixels are then removed prior to image parameterization \citep{PunchSupercuts}.  During the image parameterization stage, each Cerenkov image in the telescope's camera is characterized by an ellipse, using a moment analysis of the recorded signal amplitudes in each pixel.  Each shower is characterized by the major axis ({\em length}), minor axis ({\em width}), the angle between the major axis of an image's ellipse and a line from the centroid of the ellipse to the source position ($\alpha$), the length of this line ({\em distance}), and the overall signal of the shower ({\em size}).  These are known as Hillas parameters and they allow for the removal of the primary source of background, cosmic rays, by exploiting the intrinsic differences in the development of hadronic cosmic ray and gamma-ray showers \citep{Hillas}.  These differences result in different distributions of light at ground level and cuts can be used to select events most likely to be gamma-rays.  Standard cuts were made on the Hillas parameters, as shown in Table \ref{table:supercuts}, to reject as many background events as possible while still retaining many gamma-ray candidates.

The runs taken in ON/OFF mode obtain a direct measurement of background from the off-source runs, while in TRACKING mode a ``tracking ratio" is used to estimate the background.  Gamma-ray events should have a small $\alpha$ parameter if the source is centered in the field of view, so events with large values for $\alpha$ can be used to estimate the background if the ratio between background rates at small and large values of $\alpha$ is known in the absence of a source.  This ``tracking ratio" used here was found by taking observations of blank sky fields with no detected VHE sources throughout the 2005 -- 2006 observing season.  It is defined as the ratio of the integrated number of events between $\alpha$ = 0 -- 15$\degree$ and those between $\alpha$ = 20 -- 65$\degree$.  Once it is calculated, the tracking ratio is used to estimate background rates for all TRACKING observations.  

\subsubsection{MAGIC Data}

MAGIC is the largest single dish Cerenkov telescope in operation \citep{baixeras04,cortina}. A 17 m tesselated reflector focuses the light from air showers on a camera composed of 576 PMTs. For high elevation angle observations, the MAGIC trigger threshold currently reaches down to 50 -- 60 GeV \citep{crab-magic}.

The measurements reported in this article were conducted from 2006 April 29 21:32 UT to 2006 April 30 00:59 UT at elevation angles spanning $49^{\circ} - 81^{\circ}$. MAGIC observed the source employing the so-called \emph{wobble} mode \citep{Fomin}, during which the telescope alternates between tracking two (or more) opposing sky directions, each $0.4^{\circ}$ off-source, for 20 minutes each \citep{magic-only}. After removing events not containing sufficient information for further analysis (see Table \ref{table:MAGICcuts} top), events from accidental triggers, triggers from nearby muons, and data affected by adverse meteorological conditions, 2.8 hours out of the 3.8 hours of data were used for further analysis.

The data were processed using the analysis and reconstruction software package for MAGIC data {\citep{bretz09}}. A description of the different analysis steps can be found in \citet{gaug05}, \citet{crab-magic} and \citet{bretz05}. For this analysis, the signal has been extracted using a spline algorithm.  After calibration, the shower images were cleaned of background noise by requiring a minimum photoelectron signal in the pixels as well as temporal coincidence with adjacent pixels (time image
cleaning, see \citet{magic-only}). The recorded events are characterized
by several image parameters based on the shower light distribution
(amongst them the previously mentioned Hillas parameters) and on the
temporal shower development in the camera plane ($slope$, see
\citet{magic-only}). True gamma-ray events coming from the observed source
are extracted from the hadronic background by cuts in this image
parameter space, using the separation cuts listed at the bottom of Table \ref{table:MAGICcuts}.  Events originating from the source are selected by a cut on $\theta^2$ (where $\theta$ is the angular distance between the expected source position and the reconstructed gamma-ray arrival
direction).  Gamma-ray events are then separated from background events by a
$size$ dependent parabolic cut in $width \times length \times \pi$
\citep{riegel05}. 

The Markarian 421 observations presented here are among the first data
taken by MAGIC after major hardware updates in April 2006.  Due to these changes, a thorough examination of the data was undertaken to ensure its reliability.  Despite the hardware changes, all MAGIC
subsystems were working as expected with the exception of an unstable
trigger behavior for some PMTs, leading to an early signal arrival time
for those affected. A careful study of the systematics in the low energy region has been conducted, leading to the decision of raising the energy threshold to 250~GeV \citep{magic-only}.

\subsubsection{VERITAS Data}

{\em VERITAS} is an array of four 12 m diameter IACTs each focusing light on a camera consisting of 499 PMTs.  Utilizing multiple telescopes provides a stereoscopic view of showers, making it possible to significantly reduce muon events.  This substantially improves the low energy performance of the detector since images produced from nearby muons are otherwise virtually indistinguishable from images produced by low-energy gamma-rays.  Therefore the array allows for a lower energy threshold than possible with a single telescope of the same size operating alone.  For {\em VERITAS} this threshold is near 275 GeV.  In addition, stereoscopic observations facilitate a much improved determination of the core position of showers and thus provide improved energy resolution and background rejection.

During the 2008 observation, {\em VERITAS} took data on Markarian 421 while operating in {\it wobble} mode, with an offset angle of 0.5$\degree$.  To provide background estimations using this observation mode, we follow the reflected-region model \citep{Berge} where one collection region is placed at the source position and others of equal size are placed at equal offsets from the center of the field of view and used for background measurements.  {\em VERITAS} observed Markarian 421 for 2.5 hours on 2008 May 07 from 3:59 UT to 6:28 UT using a series of 20 minute runs, with an effective exposure time of 129.7 minutes.  The observations were taken as the source elevation ranged from 79$\degree$ to 55$\degree$.  Data quality checks confirmed good weather during the observations.
  
The {\em VERITAS} data from 2008 was analyzed with Veritas Gamma-ray Analysis Suite (VEGAS), the standard {\em VERITAS} analysis package \citep{Cogan}.  The shower images were first corrected for relative gains and cleaned to remove isolated pixels.  The images were then parameterized in a similar manner as for {\em Whipple} analysis.  Initial quality cuts were made on the images' {\em size}, pixel count and {\em distance}.  For this analysis, events that survived the initial cuts but only contained data from the two telescopes with the smallest physical separation were removed.  This smaller separation distance results in a greater number of muons surviving the initial cuts, and in addition these events can distort the calculation of impact distance due to the relatively small distance between the telescopes.  After stereoscopic reconstruction and calculation of each event's impact distance (distance from the shower core to the telescope), the events were parameterized using the quantities mean scaled width (MSW) and mean scaled length (MSL) \citep{Konopelko,Daum}.  These parameters are found by scaling the width and length parameters for a given telescope by the average expected values from simulations given an impact distance and size and then finding the average for all telescopes involved in an event.   Cuts on MSW and MSL were made to separate gamma-ray events from cosmic-ray events.  Table \ref{table:VERITAScuts} details the cuts that are discussed above.  A summary of all VHE gamma-ray data is shown in Table \ref{table:TEVresults}.

\section{Results}
	\subsection{Time Averaged VHE Gamma-Ray Spectra}
	We followed the procedures from \cite{Mohanty} in constructing a spectrum based on the {\em Whipple} data. This method depends on separate energy estimates for both on-source and off-source runs, so the runs taken in the TRACKING mode were matched to contemporaneous OFF runs taken at similar elevation angles.  The selected matched runs were taken within two days of the Markarian 421 observations and at elevation angles within 1\degree~ of the corresponding TRACKING runs.  The energy spectrum was fit using a simple power law:

\begin{equation}
\frac{\mathrm{d}N}{\mathrm{d}E} =
 F_0 \cdot 10^{-11} \cdot
 \left(\frac{E}{1\,\mathrm{TeV}}\right)^{-\alpha} \cdot
  \frac{\mathrm{photons}}{\mathrm{TeV}~\mathrm{cm}^{2}~\mathrm{s}}
\end{equation}
	
\noindent Finding best-fit parameters of $\alpha = 2.23\pm0.38$ and $F_0=2.65\pm0.77$ yielding a  $\chi^2$/d.o.f. of 0.17/3 ($P = 98.2 \%$).  Uncertainties are statistical only.

The {\em MAGIC} spectrum has been derived on the same data basis as the light curve but a somewhat looser area cut was applied. This cut yielded a constant cut efficiency as a function of energy of 90\% for Monte Carlo simulated gamma-ray events, increasing the gamma-ray event statistics at the threshold \citep{1553}. The energies of the gamma events were reconstructed using a random forest regression method \citep{rf,rf-magic} trained with Monte Carlo events.  The {\em MAGIC} spectrum was modeled similarly to the {\em Whipple} spectrum, finding best fit parameters of    $\alpha = 2.28 \pm 0.09$ and $F_0 = 2.49 \pm 0.17$, yielding a $\chi^2$/d.o.f. of 2.04/4 ($P = 72.9 \%$).  All the stated uncertainties for MAGIC are purely statistical. The energy scale is known with an uncertainty of $\pm 16 \%$, the flux normalization within a systematic error of 11\% (not including the energy scale error), and the fitted power law slope has a systematic uncertainty of $\pm 0.2$ \citep{crab-magic}.

For the MAGIC analysis, an additional systematic uncertainty is provided to account for possible effects arising from the hardware instability mentioned above. These effects consist of a moderate loss of low-energy showers as well as a minor additional uncertainty in the image parameter calculation for showers of higher energy. The effect on the differential flux level is estimated to be 10\% from 250~GeV to 400~GeV and 3\% for higher energies \citep{magic-only}.

\begin{figure}
\epsscale{0.75}
\plotone{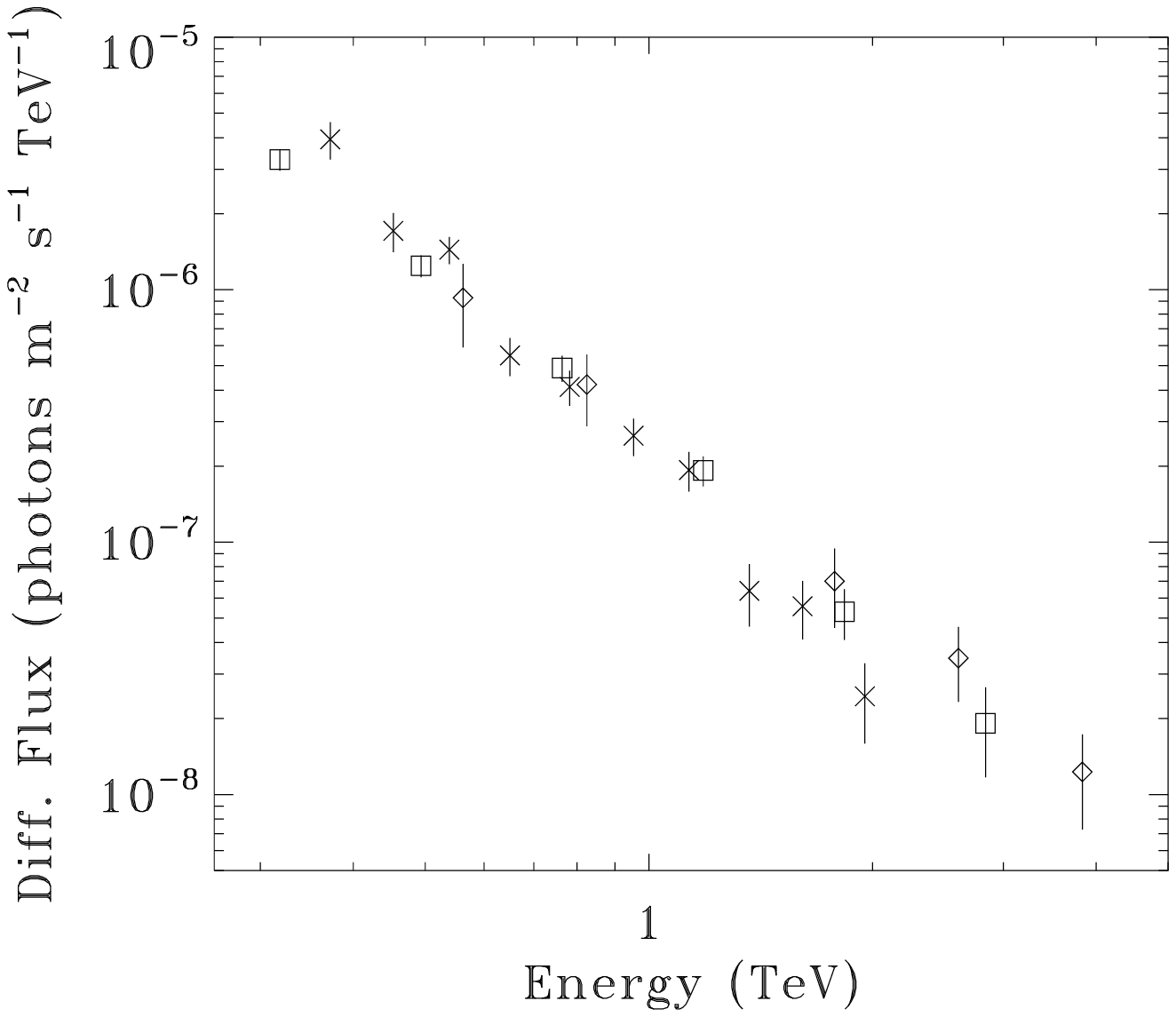}
\caption{VHE spectral analysis results, showing the data obtained with MAGIC (squares, $\alpha = 2.28 \pm 0.09$ and $F_0 = 2.49 \pm 0.17$), {\em Whipple} (diamonds, $\alpha = 2.23\pm0.38$ and $F_0=2.65\pm0.77$ ) and {\em VERITAS} (crosses, $\alpha=2.91\pm0.13$ and $F_0 = 2.01 \pm 0.15$).}
\label{fig:f1} 
\end{figure}

The spectral analysis for the {\em VERITAS} data was performed using the VEGAS analysis package \citep{Cogan}, with the same model for background estimates as used for the light curve.  The spectrum was again fit with a simple power law model, finding best fit parameters of $\alpha=2.91\pm0.13$ and $F_0 = 2.01 \pm 0.15$, yielding a $\chi^2$/d.o.f. of 9.66/8 ($P = 29.0\%$).  Again, stated uncertainties are statistical only.  The results from the spectral fits for all VHE data can be seen in Figure \ref{fig:f1}.

\subsection{Spectral Energy Distribution and Modeling}

The  X-ray spectrum was initially fit using XSPEC 12.  The data were fit with a power law modified by interstellar absorption, yielding a value for the photon index of $\alpha = 2.258\pm0.002$ and $2.153\pm0.002$ for the data obtained simultaneously with the 2006 {\em MAGIC} and {\em Whipple} 10m observations, respectively. For X-ray data taken during the 2008 {\em VERITAS} observations, a photon index of $\alpha = 2.519\pm0.010$ was found.  The hydrogen column density was left as a free parameter for all fits, finding values of $3.16\times10^{20} cm^{-2}$, $2.25\times10^{20} cm^{-2}$ and $4.51\times10^{20} cm^{-2}$ for the data obtained simultaneously with the {\em MAGIC}, {\em Whipple} 10m and {\em VERITAS} observations, respectively. Using these results, the spectrum was unfolded and de-absorbed to derive the intrinsic X-ray spectrum of Markarian 421.  Only statistical errors were taken into account here.  In addition, the count rates found for the OM exposures were converted to flux using the standard conversion factor\footnote{See:  http://heasarc.nasa.gov/docs/xmm/sas/USG/node135.html} and an average point was determined for each time interval.  In addition, using the ultraviolet extinction law from \citet{Cardelli}, the absolute extinction for the UVM2 band was calculated to be $A$(UVM2) = 0.13, allowing for de-reddening of the OM data using a correction factor of 1.13.

Figure \ref{fig:f2} shows the broadband SEDs corresponding to the three epochs of VHE observations.  It is important to note that within each epoch the multiwavelength data are genuinely simultaneous.  Spectral variability is observed between epochs.  

Modeling of the SEDs was carried out using a leptonic model \citep{Bottcher}.  In this model, the spectral distribution of injected electrons is described by a power law with low and high energy cutoffs of  $\gamma_{min}$ and $\gamma_{max}$, respectively.  The emitting region is assumed to be in a state of temporary equilibrium and of spherical shape, with radius {\em R}, and moves out along the jet at relativistic speed $v/c=(1-1/\Gamma^2)^{1/2}$, where $\Gamma$ is the bulk Lorentz factor.  As the emitting region moves along the jet, particles cool due to radiative losses and may escape from the region.  The timescale of these escapes is factored into the model as $t_{esc}=\eta R/c$, with $\eta \ge1$.  The radiative processes considered include synchrotron radiation, SSC, and inverse Compton scattering of external photons.  However, we found that a model with a negligible contribution from external photons (i.e., a pure SSC model) provides a good match to the SEDs during both observations.  In addition, the SED matches have been absorbed with the extragalactic background light model discussed in \citet{EBL}.  The values for magnetic field and the spectral index of the injected electrons were varied until a good match was found for the 2006 and 2008 SEDs.   The parameters for these models are shown in Table \ref{table:model}.  The models are also shown in Figure \ref{fig:f2}.\\

	\subsection{Cross-Band Correlation}
\begin{figure}
\epsscale{1}
\plottwo{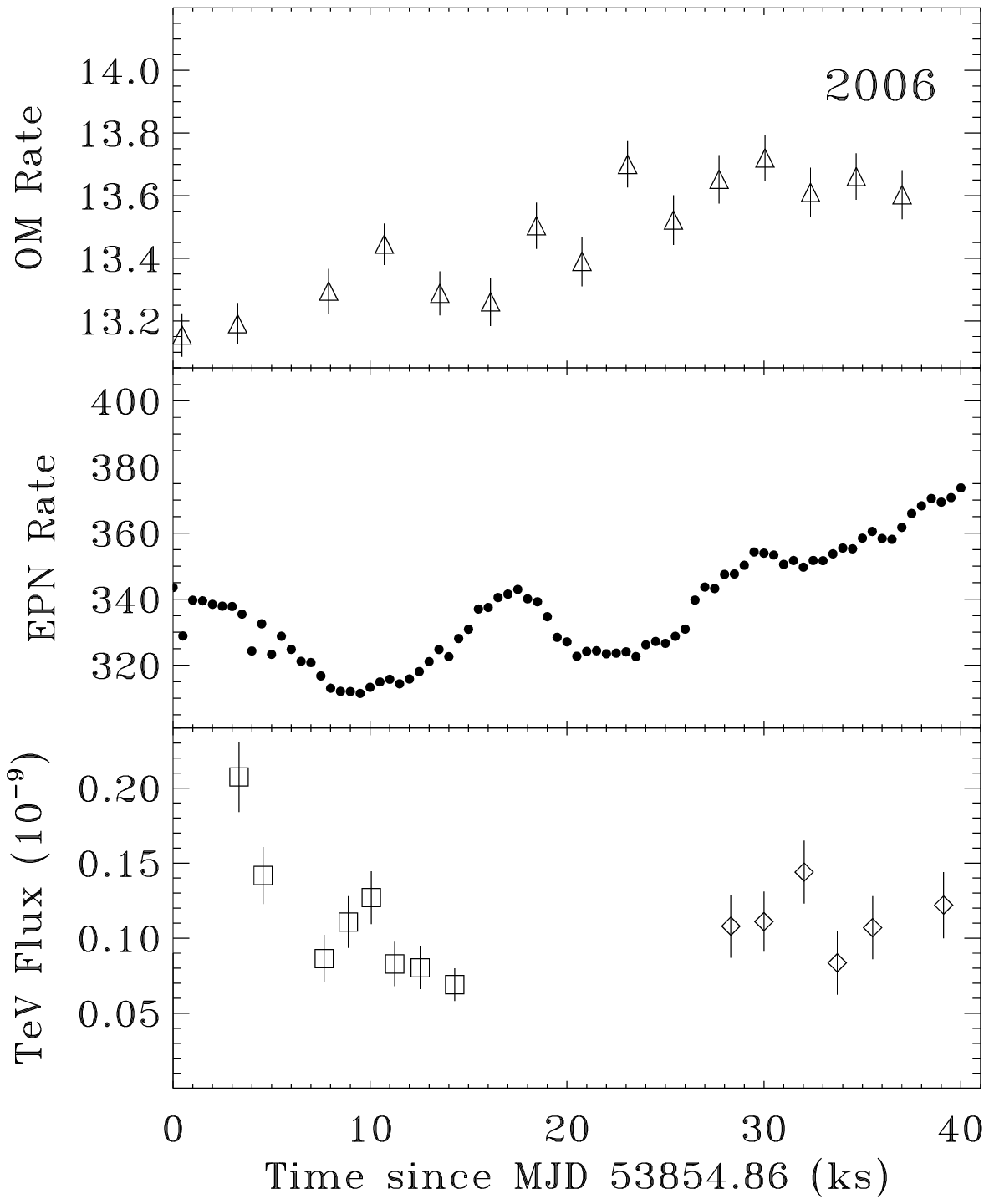}{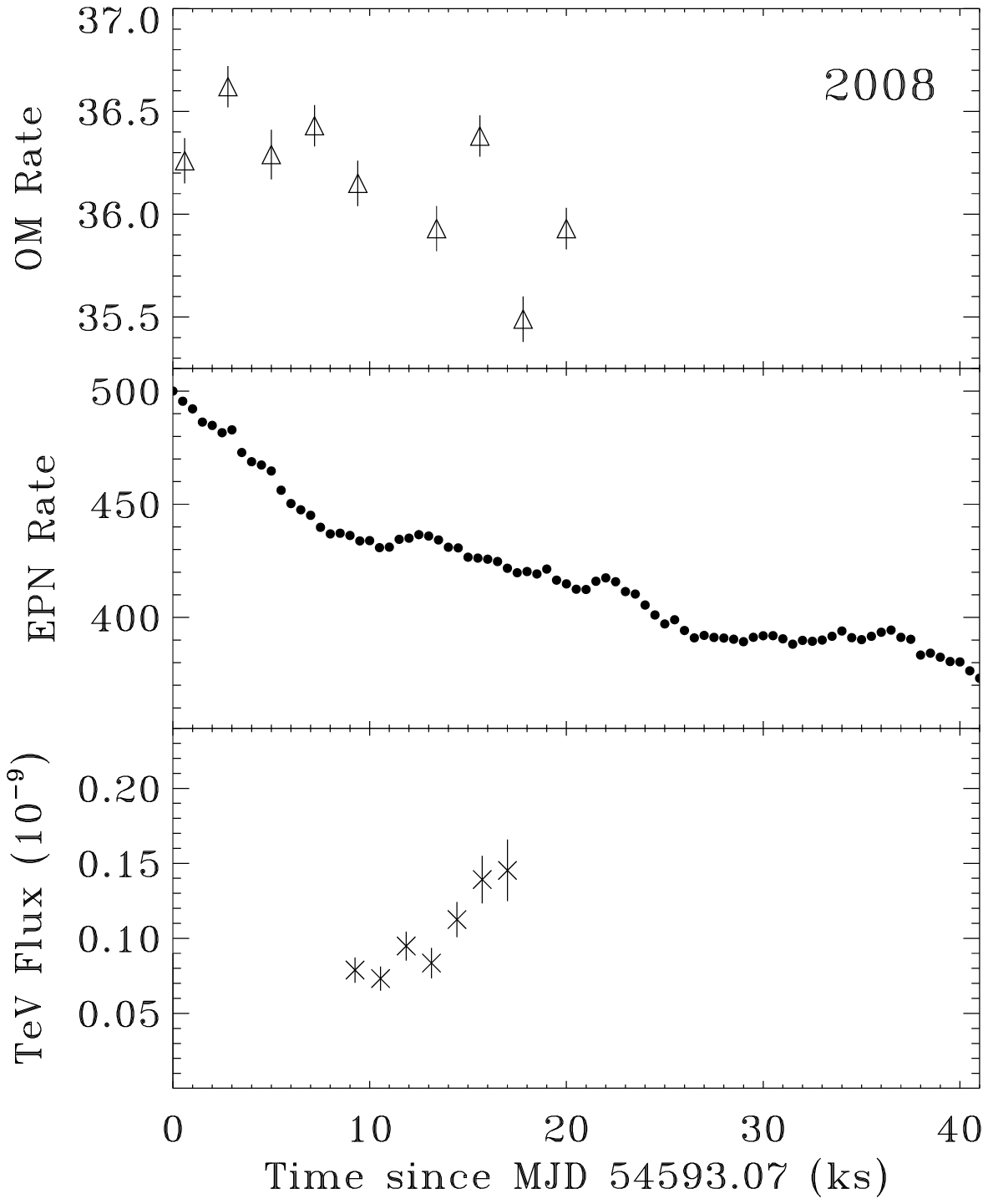}
\caption{Plot of light curves from (a) 2006 observations and (b) 2008 observations.   {\em XMM-Newton} OM data (200-300nm) is shown with triangles in units of cts s$^{-1}$.  {\em XMM-Newton} EPN data (0.5 - 10.0 keV) is shown with filled circles in units of cts s$^{-1}$.  Error bars for the EPN data are smaller than the data points.  The data from {\em MAGIC}, {\em Whipple} and {\em VERITAS} are shown with squares, diamonds, and crosses, respectively in units of photons cm$^{-1}$ s$^{-2}$ (above 250 GeV).  Note that the EPN and OM scales differ on the 2006 and 2008 panels.}
\label{fig:f3}
\end{figure}

	The VHE, X-ray and UV light curves are shown in Figure \ref{fig:f3} with 1-$\sigma$ uncertainties.  The X-ray lightcurve was initially binned using a time interval of 500 seconds, while the bins for the VHE light curves were primarily determined by the standard length of data runs for the respective telescopes.  The {\em Whipple} 10m, {\em MAGIC}, and {\em VERITAS} data are in 28-minute, 22-minute and 20-minute bins, respectively.  To provide a direct comparison between the results obtained with different VHE experiments, we show the gamma-ray fluxes above a common energy threshold (250 GeV).  To reach this threshold for the {\em Whipple} 10m, the flux was extrapolated from the power law spectrum fit to the data.
	
In addition, contemporaneous data taken on the Crab Nebula, a standard candle for VHE experiments, were studied to assess systematic uncertainty in the flux calibration of the VHE data.  The systematic uncertainty in flux calibration between the {\em MAGIC} and {\em VERITAS} experiments was estimated to be on the order of 10\% with {\em MAGIC} systematically measuring a lower flux than {\em VERITAS}.  The systematic uncertainty in the {\em Whipple} 10m and {\em VERITAS} flux calibration was found to be on the order of 40\% with the  {\em Whipple} 10m systematically measuring a higher flux than {\em VERITAS}.  These systematic uncertainties are not included in the data.

	The error bars on the VHE data are relatively large, so in order to quantify the variability, we first tested the entire light curve by fitting a constant flux value to the data.  The resulting fit had a $\chi^{2}_{\nu}$ of 3.92 for 20 degrees of freedom.  Similarly, as the MAGIC data made up the most variable portion of the VHE light curve, these data alone were fit to a constant, with a $\chi^{2}_{\nu}$ of 5.22 for 6 degrees of freedom.  This inconsistency of the data with fits to a constant indicates significant variability in the VHE band.  Markarian 421 also varied significantly at X-ray energies during the 2006 observation, with the count rate initially decreasing during the course of the {\em MAGIC} observation and slowly increasing during the {\em Whipple} observation.  Though both showed significant variability, the X-ray and VHE data from 2006 do not appear to be correlated.  	
	
	To examine the X-ray/VHE correlation more closely, we show in Figure \ref{fig:f4} the measured VHE flux and X-ray count rates.  The X-ray data were rebinned to match the resolution of the corresponding VHE data.  To quantify the correlation and examine the uncertainty in this correlation, we used a method similar to that discussed in \citet{Albert07}.  A set of 25000 light curves was simulated based on the Gaussian errors of the data points in the X-ray and VHE bands, respectively.  For each pair of simulated X-ray/VHE light curves, the value of Pearson's r was calculated.  A histogram of these possible r values was generated, giving an average r value of $-0.050 \pm 0.050$, indicating a lack of correlation.  

	To compare to previous work, the best fit correlation between the X-ray and VHE bands found in \citet{Blazejowski} is plotted in Figure \ref{fig:f4} (dotted line).  In order to carefully compare this previous result to our results, the PIMMS tool\footnote{See:  http://heasarc.gsfc.nasa.gov/Tools/w3pimms.html} was used to convert the {\em RXTE} count rates to {\em XMM-Newton} count rates, taking into account energy range, spectral shape and hydrogen column density.  To more directly compare our data to \citet{Blazejowski}, we scaled the correlation to match our {\em Whipple} 10m points in Figure \ref{fig:f4} (dot-dashed line).  The overall normalization shift could be a reflection of hysteresis on long timescales.  Although the scatter in the VHE points are large, the {\em VERITAS} points are systematically below the scaled correlation.  This effect is made worse by the fact that the {\em Whipple} 10m points are known to be systematically higher than the {\em VERITAS} points (by 40\%).


There appears to be no correlation between the VHE and UV variations, most notably during the first half of the 2006 observation, where the UV rates increase as the VHE flux decreases.
However, the UV rates during both observations appear to roughly follow the trend of the X-ray rates, with a significantly higher rate during the 2008 observation.  Using the same method as used above to examine correlations between the VHE and X-ray bands, we found an average value for Pearson's r of $0.940 \pm 0.001$ for the total set of X-ray and UV data, indicating strong correlation.  
	
\subsection{Spectral Hysteresis} \label{sec:hysteresis}

\begin{figure}
\epsscale{0.6}
\plotone{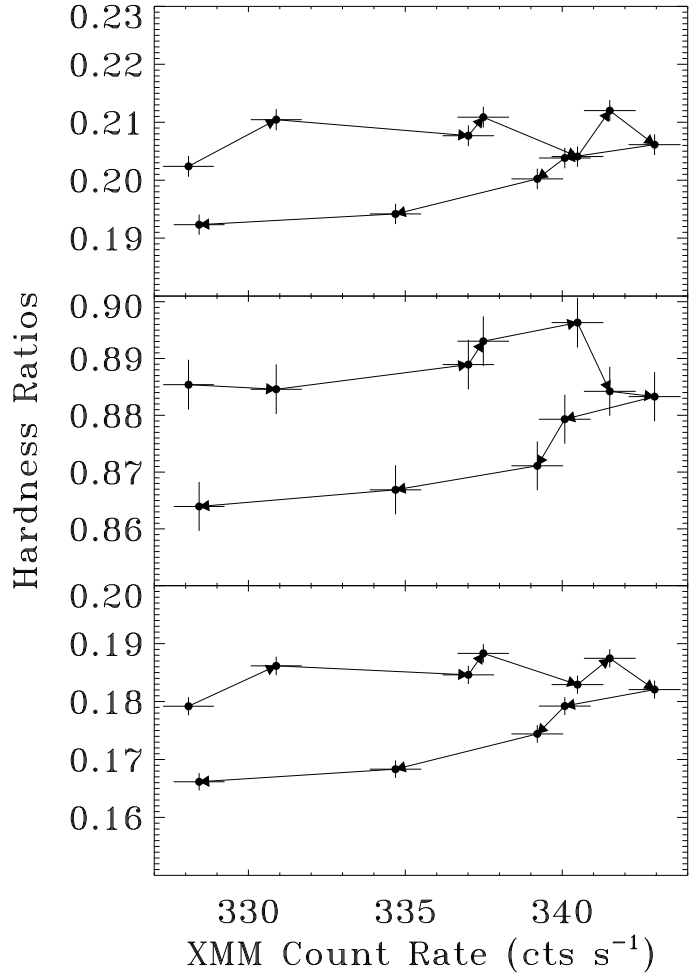}
\caption{{Hardness ratio vs intensity plots to study spectral hysteresis during the increase in activity centered around 17 ks into the 2006 {\em XMM-Newton} observation.  Three bands are used for this analysis.  The upper panel uses (3-10 keV)/(1-3 keV) for the hardness ratio, the middle panel uses (1-3 keV)/(0.5-1 keV), and the bottom panel uses (3-10 keV)/(0.5-1 keV).  The arrows indicate the progression of time.  Clockwise hysteresis is present in all three panels, indicating a lag in the lower energy bands.}}
\label{fig:f5}
\end{figure}

We also examined the spectral evolution of the X-ray data to determine if any hysteresis was present during the observations indicating a dependence of the system on previous states.  The data were divided into three energy bands in order to calculate hardness ratios:  0.5 - 1 keV, 1 - 3 keV, and 3 - 10 keV.  These bands were chosen such that each band contains a roughly equal number of counts.  Figure \ref{fig:f5} shows the evolution of the hardness ratios versus intensity (in terms of count rates) through the flare seen in the 2006 data (which peaks near 17-18 ks into the observation; see Figure \ref{fig:f3}).  Spectral hysteresis is clearly seen during this rapid flare.  Clockwise patterns indicate a lag in the response of lower-energy photons with respect to that of higher-energy photons.  Similar patterns have been seen previously \citep[e.g.,][]{Brinkmann}, though counter-clockwise patterns have been observed as well \citep{Cui}.  Different hysteresis patterns indicate a complex mechanism that may differ between outbursts.

No rapid flares were observed during the 2008 observation.  Examination of the entire 2008 observation did not reveal any overall spectral hysteresis pattern occurring on the order of several hours.  Similarly, no overall hysteresis was found when considering the entire 2006 observation.

\section{Discussion}

This work provides the results from a ToO campaign with the primary focus of studying the rapid flaring activity of blazars on sub-hour timescales in both the X-ray and VHE bands.  In addition, these observations expand the pool of truly simultaneous, multiwavelength observations of blazars, essential for the study of correlations during rapid flares.  The variability observed during the 2006 outburst in the simultaneous X-ray and VHE observations should be enough to provide useful information about a correlation, if any, between the two energy bands.  In addition, the observations from 2008 during a separate outburst, provide an extended picture for investigating possible correlations between these two bands.  Rapid X-ray and VHE flares are often expected to be correlated in one-zone SSC scenarios, as all photons are expected to originate from the same population of electrons.  Examination of the observed SEDs can provide additional constraints on model parameters and reveal what factors strongly affect each flare's spectral profile.  

\begin{figure}
\epsscale{0.75}
\plotone{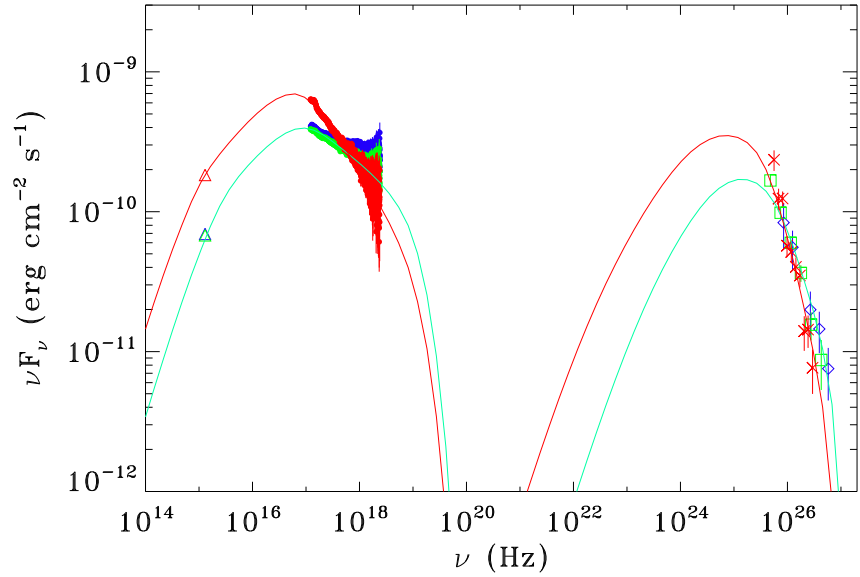}
\caption{{Spectral energy distribution with the SSC model for the 2006 (teal blue) and 2008 (red) data.  The data from {\em XMM-Newton OM}, {\em XMM-Newton EPN}, {\em MAGIC}, {\em Whipple} and {\em VERITAS} are shown with triangles, filled circles, squares, diamonds, and crosses, respectively, with data taken during the {\em MAGIC}, {\em Whipple} and {\em VERITAS} observation times shown in green, blue and red.}}
\label{fig:f2}
\end{figure}

The SED fits show a steeper spectrum for the 2008 observation than for the 2006 observation, particularly at X-ray energies.  This is intriguing as previous investigations have observed spectral hardening with increased flux \citep[e.g.,][] {Xue}.  In addition, both peaks in the SED show a slight shift to lower energies during the 2008 observation where previous investigation has shown a shift to higher energies with increased luminosity \citep[e.g.,][]{Blazejowski}.  Figure \ref{fig:f2} shows that our one-zone SSC model fits both the 2006 and 2008 data quite well.  

\begin{figure}
\epsscale{1}
\plotone{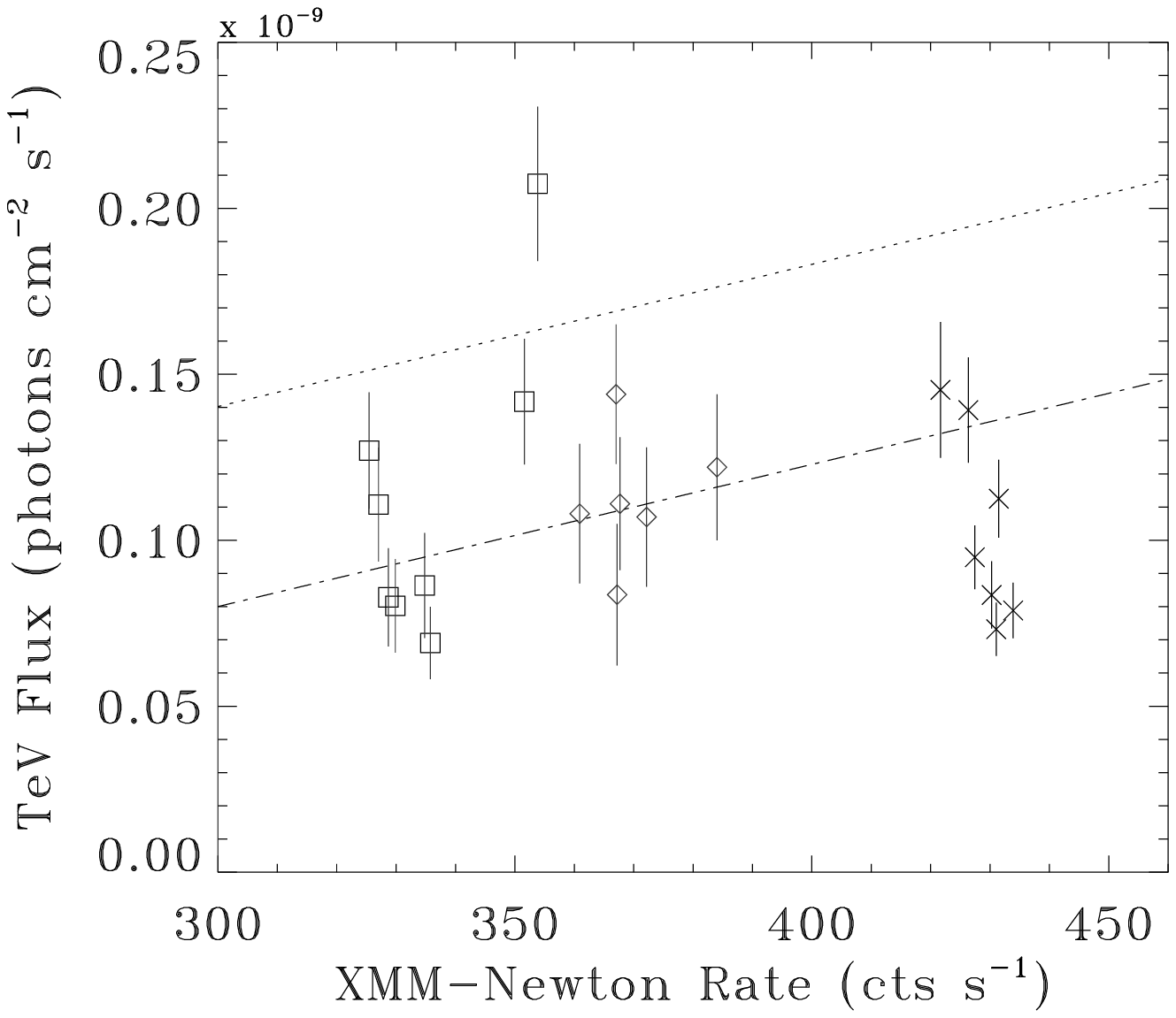}
\caption{{Plot of simultaneous {\em XMM-Newton} EPN (0.5 - 10 keV) and VHE data to search for correlation between the two bands.  MAGIC data points are designated by squares, {\em Whipple} points by diamonds, and {\em VERITAS} points by crosses.  All {\em XMM-Newton} data used for this plot was analyzed with XMMSAS version 8.0.0 to maintain consistency.  The dotted line is the X-ray/VHE correlation best fit from \citet{Blazejowski} converted to the appropriate units.  The dot-dashed line has the same slope but is scaled to the average of our {\em Whipple} 10m results.}}
\label{fig:f4}
\end{figure}

Historically, multiwavelength monitoring campaigns have observed correlation between the X-ray and VHE emission in Markarian 421, though the data sets are not comprised entirely of strictly simultaneous multiwavelength data \citep[e.g.,][]{Blazejowski,Fossati}.  Though the dynamical range of the observed variability is not as large as that seen in previous long-term monitoring campaigns, we seem to be seeing short-term variability that shows a different correlation pattern here.  Surprisingly, both the 2006 data alone and the entire 2006/2008 data set show no obvious correlation, with a Pearson's $r$ for the entire X-ray/VHE data set of $-0.050 \pm 0.050$, though the SSC model fits the SEDs well (see Figure \ref{fig:f2} and Figure \ref{fig:f4}).  This implies that the X-ray and VHE photons may originate from electrons with different energies, similar to what was observed during a study of PKS2155-304 \citep{pks2155}.  Other scenarios that could explain the observed variability patterns include the possibility of an inhomogeneous emission region or hadronic origin of the VHE emission.  It is interesting to note that Markarian 421 is clearly behaving very differently here from what is usually reported of the source \citep[e.g.,][]{Blazejowski}, where the X-ray and VHE variabilities are seen to be strongly correlated.  Such personality makes it difficult to generalize the results to other blazars.

In addition, the OM allowed for a detailed search for UV/X-ray or UV/VHE correlations.  Though previous multiwavelength campaigns on Markarian 421 have obtained optical, X-ray and VHE data, correlations between the X-ray and optical data have either not been studied or were not significant \citep[e.g.,][]{Albert421, Blazejowski, Horan}.  This study found the average rate from the {\em XMM-Newton} PN detector during the 2008 observation was about 20\% higher than the rates measured during the 2006 observation.  Examination of the OM data in the UV band shows that the count rate more than doubles between the 2006 and 2008 observation.  The UV rate also appears to vary in step with the X-ray rate within the individual observations.  Calculation of the Pearson's r value for the total UV/X-ray data set yielded a value of $0.940 \pm 0.004$, indicating a strong correlation.  This provides direct observational evidence for a link between the emission mechanisms at X-ray and UV wavelengths.  On the other hand, we found no apparent UV/VHE or X-ray/VHE correlation, implying that the seed photons for the inverse Compton process (to produce the VHE photons) cannot be provided by the UV or X-ray emission observed.

We also searched for hysteresis patterns in the X-ray data to provide further information about the outbursts.  Hysteresis was observed during one rapid flare in the X-ray data from the 2006 observation.  No rapid flares occurred during the 2008 observation.  Spectral hysteresis has been commonly observed in blazars, but the phenomenon is not yet understood completely.  In \citet{Kirk}, a simple model is used to produce spectral hysteresis patterns that may be observed during the course of a flare.  In this model, the behavior is characterized by the relationship between three timescales associated with the duration of the flare variability ($t_{var}$), synchrotron cooling ($t_{cool}$) and particle acceleration ($t_{acc}$).  The relationship between these timescales result in four possible cases that are discussed in detail.  The clockwise hysteresis found in the X-ray data, indicating a lag at lower energies in the X-ray band, coupled with the essentially symmetric shape of the flare in the X-ray light curve seems to indicate that the case with $t_{cool} \gg t_{var} \gg t_{acc}$ is most relevant to this observation.  The lag in the low energy photons is a result of the inverse relationship between $t_{cool}$ and the energy of the cooling particles.  Although we observed clockwise patters in one of the observations, other patterns have also been observed.  This is yet another example of the personality of the source and may indicate physical differences between individual flares and outbursts.

{\it Acknowledgments.  }The {\em VERITAS} research was supported by grants from the U.S. Department of Energy, the U.S. National Science Foundation and the Smithsonian Institution, by NSERC in Canada, by Science Foundation Ireland and by STFC in the U.K..  We acknowledge the excellent work of the technical support staff at the FLWO and the collaborating institutions in the construction and operation of the instrument.  D.G. and W.C. wish to acknowledge support by NASA through grants NNX06AB96G, NNX08AD76G, and NNX08AX53G.  The {\em MAGIC} collaboration thanks the Instituto de Astrofisica de Canarias for
the excellent working conditions at the Observatorio del
Roque de los Muchachos in La Palma. The support of
the German BMBF and MPG, the Italian INFN, and
Spanish MCINN is gratefully acknowledged. This work
was also supported by ETH Research Grant TH 34/043,
by the Polish MNiSzW Grant N N203 390834, and by
the YIP of the Helmholtz Gemeinschaft.

{\it Facilities:} \facility{MAGIC}, \facility{VERITAS}, \facility{FLWO:10m}, \facility{XMM}

\begin{table}[h]
\caption{Summary of Markarian 421 {\em XMM-Newton} Observations.}
\label{table:xmmobs}      
\centering
\begin{tabular}{ccccc}
\hline\hline                             
Obs. ID     &   Start Time                & PN Exposure & Avg. PN Rate&     Avg. OM Rate \\   	 
            &               &   (ks)       &      (cts s$^{-1}$) & 	   	(cts s$^{-1}$)  \\	    
           \hline	 
 0302180101 &  2006 Apr 29           &   41.9   & 336.30 $\pm$ 0.09  & 13.5$\pm$0.2 \\       
            &  20:44UT           &                       &                &         \\  
0502030101 & 2008 May 07 & 43.2    &        411.31 $\pm$ 0.12 & 35.82 $\pm$ 0.03\\
       &  01:34UT           &          &                           &         \\  
\hline
\end{tabular}
\end{table}

\begin{table}[htdp]
\caption{Whipple Event Selection Parameters}
\begin{center}
\begin{tabular}{l c}
\hline \hline
Parameter & Values for Cuts \\ \hline
Trigger level& $>$ 30 digital counts \\ 
Shape cuts& 0.05\degree $<$ width $<$ 0.12\degree \\ 
& 0.13\degree $<$ length $<$ 0.25\degree \\ 
Muon cut & length/size $<$ 0.0004\degree dc$^{-1}$ \\ 
Distance cut & 0.4\degree $<$ distance $<$ 1.0\degree\\ 
Orientation cut & $\alpha$ $<$ 15\degree \\ \hline
\end{tabular}
\end{center}
\label{table:supercuts}
\end{table}

\begin{table}[htdp]
\caption{MAGIC Cuts.}
\tablecomments{{L}eakage parameter is defined as the ratio of the signal in the outer pixels of the camera to the total signal.}
\begin{center}
\begin{tabular}{l c}
\hline \hline
Parameter & Values for Cuts \\ \hline
Number of shower islands & $<$  3\\ 
Number of used pixels & $>$ 5\\
Leakage & $<$ 0.3\\ \hline
Gamma/Hadron Separation Cuts:\\
$\theta^2$& $<$ 0.046  \\ 
Area &  $< 0.265\times(1 - 0.0803\times(log10(size)-5.77)^2)$ \\ 
Standard time cuts & $slope > (distance-0.5)\times7.2$\\ \hline
\end{tabular}
\end{center}
\label{table:MAGICcuts}
\end{table}

\begin{table}[htdp]
\caption{VERITAS Cuts}
\begin{center}
\begin{tabular}{l c}
\hline \hline
Parameter & Values for Cuts \\ \hline
{\em Size} & $>$  400 digital counts\\ 
Pixels & $\geq$ 5\\
{\em Distance} & $\leq$ 1.43$\degree$ \\ \hline
Mean Scaled Width & 0.05 $<$ MSW $<$ 1.16 \\ 
Mean Scaled Length & 0.05 $<$ MSL $<$ 1.36 \\ \hline
\end{tabular}
\end{center}
\label{table:VERITAScuts}
\end{table}

\begin{table}[htdp]
\caption{Gamma-ray data run information.}
\tablecomments{Some data were removed due to adverse meteorological conditions.}
\begin{center}
\begin{tabular}{c c c c c c}
\hline \hline
Facility/Date &Time& Wobble Direction & Significance ($\sigma$) \\ \hline 
MAGIC\\
2006 Apr 29 &21:29 UT &  0.40$\degree$+000&  12.0\\    
&21:49 UT  &0.40$\degree$+180&   9.3\\
& 22:41 UT & 0.40$\degree$+180 &  5.0 \\
& 22:51 UT & 0.40$\degree$+000   &7.6 \\
  & 23:11 UT  &0.40$\degree$+180  & 8.4\\
 & 23:31 UT&0.40$\degree$+000  & 8.9  \\
&	 23:52 UT&  0.40$\degree$+180  & 6.6 \\
  2006 Apr 30& 00:12 UT & 0.40$\degree$+000 &  6.0 \\
& 00:32 UT & 0.40$\degree$+180  & 5.9   \\
& 00:53 UT& 0.40$\degree$+000  & 4.4  \\ 
&&TOTAL&23.7\\ \hline
WHIPPLE\\
2006 Apr 30&04:37 UT& TRACKING &4.9\\
&05:05 UT&TRACKING &5.4\\
&05:39 UT& TRACKING&6.6\\
&06:07 UT& TRACKING&3.7\\
&06:37 UT& ON/OFF&4.8\\
&07:37 UT& ON/OFF&5.3\\
&&TOTAL&12.5\\ \hline
VERITAS\\ 
2008 May 07&03:59 UT&0.5$\degree$ N &15.5\\
&04:21 UT&0.5$\degree$ S &15.4\\
&04:43 UT& 0.5$\degree$ E &16.9\\
&05:04 UT& 0.5$\degree$ W &13.5\\
&05:26 UT& 0.5$\degree$ N&15.2\\
&05:47 UT& 0.5$\degree$ S &14.6\\
&06:08 UT& 0.5$\degree$ E &13.7\\ 
&&TOTAL&39.6\\ \hline
\end{tabular}
\end{center}
\label{table:TEVresults}
\end{table}

\begin{table}[htdp]
\caption{Model Parameters}
\begin{center}
\begin{tabular}{l c c}
\hline \hline
Parameter & 2006 Value&2008 Value\\ \hline
$\gamma_{min}$:            					&$4.2\times10^{4}$	&$3.3\times10^{4}$\\
$\gamma_{max}$:                       				&$5.0\times10^{5}$	&$4.0\times10^{5}$\\
injection electron spectral index:        		&2.6		                   &3.2\\
Escape time parameter ($t_{esc} = \eta R/c)$	&$\eta_{esc}$ = 3	&$\eta_{esc}$ = 3\\
Magnetic field at $z_{0} $[G]:                    		&0.48			&0.68\\
Bulk Lorentz factor:                 				&$\Gamma $= 20	&$\Gamma $= 20\\
Blob radius [cm]                    		&$3.0\times10^{15}$ &$3.0\times10^{15}$\\ 
Observing angle [degrees]				&$\theta_{obs} = 2.87$&$\theta_{obs} = 2.87$\\
$L_e$ (jet)		&$7.76\times10^{42}$ erg s$^{-1}$&$1.06\times10^{43}$ erg s$^{-1}$\\
$L_B$ (jet)		&$3.11\times10^{42}$ erg s$^{-1}$&$6.24\times10^{42}$ erg s$^{-1}$\\
$L_B/L_e$		&0.40&0.59\\ \hline
\end{tabular}
\end{center}
\label{table:model}
\end{table}

\end{document}